\documentclass[12pt,a4paper]{article}
\pdfoutput=1

\usepackage{bm}
\usepackage{amssymb}
\usepackage{amsmath}
\usepackage{amsfonts}
\usepackage{authblk}
\usepackage{setspace}
\usepackage[pdftex]{graphicx} 

\numberwithin{equation}{section}

\newcommand{\mtilde}{\widetilde{m}}

\newcommand{\Mhat}{\widehat{M}}

\newcommand{\BbbR}{\mathbb{R}}

\DeclareMathOperator{\Realpart}{Re}

\DeclareMathOperator{\Tr}{Tr}

\DeclareMathOperator{\Ci}{Ci}

\title{Unruh-DeWitt detector response\\ 
across a Rindler firewall is finite}
\author{Jorma Louko} 
\affil{School of Mathematical Sciences\\ 
University of Nottingham\\ 
Nottingham NG7 2RD\\ 
UK}

\date{{\small July 2014, revised August 2015}\\[1ex]
{\small Published in JHEP {\bf 1409} (2014) 142}\footnote{This 
eprint differs from the JHEP version in 
the second full paragraph after equations~\eqref{eq:DeltaFboth}, 
correcting 
the parity description of~$\Delta \mathcal{F}^{(0)}$.}}

\begin{document}

\maketitle

\begin{abstract}
We investigate a two-level Unruh-DeWitt 
detector coupled to a massless scalar field or 
its proper time derivative in $(1+1)$-dimensional Minkowski spacetime, 
in a quantum state whose correlation structure across 
the Rindler horizon mimics the stationary aspects of a firewall 
that Almheiri \emph{et al\/} 
have argued to ensue in an evaporating black hole spacetime. 
Within first-order perturbation theory, we show that the detector's 
response on falling through the horizon is sudden but finite. 
The difference from the Minkowski vacuum response is proportional to 
$\omega^{-2}\ln(|\omega|)$ for the non-derivative 
detector and to $\ln(|\omega|)$ for the derivative-coupling 
detector, both in the limit of a large energy gap 
$\omega$ and in the limit of adiabatic switching.  
Adding to the quantum state high Rindler temperature excitations behind the horizon 
increases the detector's response proportionally to the temperature; 
this situation has been suggested to model the energetic curtain proposal of 
Braunstein \emph{et al\/}. 
We speculate that the $(1+1)$-dimensional derivative-coupling detector 
may be a good model for a non-derivative detector that crosses a firewall
in $3+1$ dimensions. 
\end{abstract}


\newpage 

\section{Introduction\label{sec:intro}}

If black hole evaporation is assumed to preserve unitarity, a range of
arguments based on quantum correlations
\cite{braunstein-et-al,Mathur:2009hf,Almheiri:2012rt} suggest that
physics at the slowly-shrinking horizon may differ significantly from
the innocuous picture that
underlies Hawking's original derivation of black hole
radiation within curved spacetime quantum field theory~\cite{hawking}. 
In particular, Almheiri \emph{et al\/}
\cite{Almheiri:2012rt} have argued that the horizon will be replaced
by a region of high curvature, a ``firewall'', which will destroy any
observer who attempts to fall into the black hole.  Reviews with
extensive references can be found in
\cite{Susskind:2013tg,Almheiri:2013hfa,Page:2013mqa}.

A key element in the firewall argument as formulated in \cite{Almheiri:2012rt} 
is that the conventional quantum field theory picture of black 
hole evaporation involves strong
quantum correlations between the black hole interior and exterior, and
the assumption of unitary evolution of the full system turns out to
preclude such correlations. In this paper we examine the consequences
of severing closely similar correlations across a Killing horizon in a
system in which the requisite quantum state can be readily written
down: a conformal scalar field in $1+1$ spacetime dimensions. For
concreteness, we take the spacetime to be Minkowski, so that the sense
of thermality is that of the Unruh effect of uniform
acceleration~\cite{unruh,Crispino:2007eb}, and we induce a firewall
by breaking the correlations across the Rindler horizon.
The Killing horizon in $(1+1)$-dimensional black hole spacetimes with
a Kruskal-like global structure could be treated in the same manner,
with similar conclusions.

We shall not attempt to examine how the spacetime geometry might react
to the firewall singularity of the scalar field on the Rindler
horizon, but we shall examine how the singularity 
of the scalar field affects a particle
detector that falls through the horizon. We consider 
a two-level Unruh-DeWitt (UDW) detector that couples 
linearly to the scalar field
\cite{unruh,dewitt,birrell-davies,wald-smallbook}, 
and its modification that couples linearly to the proper time derivative of the field 
\cite{Raine:1991kc,Raval:1995mb,Davies:2002bg,Wang:2013lex,Martin-Martinez:2014qda,Juarez-Aubry:2014jba}. 
The reasons to consider the derivative-coupling detector are twofold. 
First, for quantum
states that are regular in the Hadamard sense~\cite{decanini-folacci},
the derivative-coupling detector is insensitive to the infrared
ambiguity in the Wightman function of the $(1+1)$-dimensional
conformal field. Second, the short-distance behaviour of the
$(1+1)$-dimensional derivative-coupling UDW detector is similar to
that of the $(3+1)$-dimensional UDW detector with a non-derivative
coupling~\cite{Juarez-Aubry:2014jba,satz,louko-satz-curved,hodgkinson-louko}.
We may hence expect a derivative-coupling detector in $1+1$ dimensions
to be a good model for a non-derivative detector that crosses a
$(3+1)$-dimensional firewall. 
We recall that the non-derivative UDW detector in $(3+1)$ dimensions 
models the $\bm p \, \cdot \bm A$ term by which an atomic electron couples to the
quantised electromagnetic field when there is no angular momentum
exchange \cite{MartinMartinez:2012th,alhambra}. 

We shall show that crossing the Rindler firewall has a
nonzero and sudden but \emph{finite\/} effect on the detector's
transition probability, within first-order perturbation theory. 
In terms of the detector's energy gap~$\omega$, 
the difference from the Minkowski vacuum transition probability 
is proportional to $\omega^{-2}\ln(|\omega|)$ 
for the non-derivative detector and to $\ln(|\omega|)$ 
for the derivative-coupling detector both in the limit 
of a large energy gap and in the limit of adiabatic switching. 

We consider also a generalisation to a quantum state in 
which Rindler excitations are added behind the Rindler horizon 
in a way that has been 
suggested \cite{braunstein-private} to model the ``energetic curtain'' of 
\cite{braunstein-et-al} in a black hole spacetime. We show 
that in this state the response across the horizon
is again finite but can be made arbitrarily large by increasing the
temperature parameter that characterises 
the added excitations. 

We begin by reviewing in Sections \ref{sec:detector} and
\ref{sec:minkowski} the two-level UDW detector and 
its derivative-coupling generalisation, 
coupled to a massless scalar field in $(1+1)$-dimensional Minkowski spacetime. 
The Rindler firewall quantum state is constructed in 
Section \ref{sec:firewallstate-def}, and we 
discuss the sense in which it models the stationary aspects of the 
black hole firewall of~\cite{Almheiri:2012rt}. 
The response of an inertial detector that crosses the Rindler horizon 
in this state is analysed in Section~\ref{sec:resp-across}, 
deferring technical steps to two appendices.
Section \ref{sec:curtain} addresses the generalisation 
to a state in which excitations have been added behind the Rindler horizon. 
Section \ref{sec:conc} presents a summary and concluding
remarks, including a discussion of detectors with multiple levels.

We use metric signature $(-+)$ in which a timelike vector has negative
norm squared, and we set 
$c=\hbar=1$. Spacetime points are denoted by Sans Serif letters 
($\mathsf{x}$)
and complex conjugation is denoted by an overline. 


\section{Two-level UDW detector\label{sec:detector}}

We consider a pointlike two-state UDW detector, 
moving in a relativistic spacetime 
on a smooth timelike worldline $\mathsf{x}(\tau)$ parametrised by the
proper time~$\tau$. The detector's orthonormal energy eigenstates are
$|0\rangle_D$ and $|\omega\rangle_D$, with the respective
eigenenergies $0$ and~$\omega$, where $\omega$ is a real-valued
parameter. $|0\rangle_D$ is the ground state when $\omega>0$ and the
excited state when $\omega<0$. We refer to the detector as a two-level
detector. The analysis will cover also the special case $\omega=0$
in which the two states are degenerate in their energy. 
We start with arbitrary spacetime dimension but 
will shortly specify to $1+1$. 

We couple the detector to a real scalar field $\phi$ via the interaction
Hamiltonian
\begin{align}
H^{(p)}_{\text{int}}
&=
c\chi(\tau)\mu(\tau) \, \frac{d^p}{d\tau^p} \phi\bigl(\mathsf{x}(\tau)\bigr) 
\ , 
\label{eq:Hint-derivative}
\end{align}
where $c$ is a coupling constant, $\mu$ is the detector's monopole
moment operator, the parameter $p$ is a non-negative integer, 
and the switching function $\chi$ specifies how the
interaction is switched on an off. 
We assume $\chi$ to be take non-negative real values and to be 
smooth with compact support. 
For $p=0$ the detector couples 
to the value of the field at the 
detector's location, 
and for 
$p>0$ the detector couples to the $p$th-order proper 
time derivative of the field at the detector's location. 
For the reasons discussed in Section \ref{sec:intro} 
we shall mainly be interested 
in the cases $p=0$, 
which is the usual UDW detector \cite{unruh,dewitt,birrell-davies,wald-smallbook}, 
and $p=1$ 
\cite{Raine:1991kc,Raval:1995mb,Davies:2002bg,Wang:2013lex,Martin-Martinez:2014qda,Juarez-Aubry:2014jba}, 
but we shall keep the value of 
$p$ general until it needs to be specified. 

Taking the detector to be initially in the state $|0\rangle_D$ and the
field to be in a (for the moment pure) state~$|\psi\rangle$, and working in
first-order perturbation theory in~$c$, the probability for the
detector to have made a transition to the state $|\omega\rangle_D$
after the interaction has ceased can be written for all 
$p$ by a straightforward adaptation of the $p=0$ 
analysis \cite{unruh,dewitt,birrell-davies,wald-smallbook}. The outcome is 
\begin{align}
\label{eq:prob}
P^{(p)}(\omega)=c^2{|_D\langle0|\mu(0)|\omega\rangle_D|}^2\mathcal{F}^{(p)}(\omega)
\ , 
\end{align}
where the response function $\mathcal{F}^{(p)}$ is given by 
\begin{align}
\label{eq:respfunc-def}
\mathcal{F}^{(p)}(\omega)
&=
\int^{\infty}_{-\infty}\,d\tau'\,\int^{\infty}_{-\infty}\,d\tau''\, 
e^{-i \omega(\tau'-\tau'')} \,\chi(\tau')\chi(\tau'') \, 
\partial^p_{\tau'}
\partial^p_{\tau''}
\mathcal{W}(\tau',\tau'')
\ , 
\end{align}
and the correlation function $\mathcal{W}$ is the pull-back of the Wightman 
function in the state $|\psi\rangle$ to the detector's worldline, 
\begin{align}
\mathcal{W}(\tau',\tau'') :=
\langle\psi|\phi\bigl(\mathsf{x}(\tau')\bigr)
\phi\bigl(\mathsf{x}(\tau'')\bigr)|\psi\rangle
\ . 
\label{eq:W-def}
\end{align}
The integrals in \eqref{eq:respfunc-def} are
understood in the distributional sense, and they are well defined whenever
$|\psi\rangle$ is Hadamard
\cite{Fewster:1999gj,junker,hormander-vol1,hormander-paper1}, 
which we shall assume until this needs to be relaxed in 
Sections \ref{sec:resp-across} and~\ref{sec:curtain}. 
For mixed states \eqref{eq:W-def} is replaced by the 
pull-back of the mixed state Wightman function. 
From now on we drop the factor
$c^2{|_D\langle0|\mu(0)|\omega\rangle_D|}^2$ and refer to
$\mathcal{F}^{(p)}$ as the transition probability, 
or as the response. 

We now specialise to two spacetime dimensions. 
Using 
$\mathcal{W}(\tau',\tau'') = \overline{\mathcal{W}(\tau'',\tau')}$, 
we may write $\mathcal{F}^{(0)}$ as 
\begin{align}
\mathcal{F}^{(0)}(\omega)
&=
2  \int_{-\infty}^{\infty} \! d u \, \int_{0}^{\infty} 
\! ds \,  
\chi(u) \chi(u-s) \Realpart \left[ e^{-i \omega s} \mathcal{W}(u,u-s) \right] 
\ , 
\label{eq:F0-final-nice}
\end{align}
where $s=0$ does not require distributional treatment 
since 
in two dimensions the short distance singularity of the Wightman function
is merely logarithmic \cite{decanini-folacci} and hence integrable.  
A~corresponding expression for $\mathcal{F}^{(1)}$ is \cite{Juarez-Aubry:2014jba} 
\begin{align}
\mathcal{F}^{(1)}(\omega)
&=
-\omega \Theta(-\omega) 
\int_{-\infty}^{\infty} d u \, {[\chi(u)]}^2
+ 
\frac{1}{\pi}
\int^{\infty}_{0} 
ds \, 
\frac{\cos(\omega s)}{s^2} 
\int_{-\infty}^{\infty} d u \, 
\chi(u) [\chi(u) - \chi(u-s)] 
\notag
\\[1ex]
&\hspace{3ex}
+ 2  \int_{-\infty}^{\infty} \! d u \, \int_{0}^{\infty} 
\! ds \,  
\chi(u) \chi(u-s) \Realpart \left[ e^{-i \omega s}
\left(\mathcal{A}(u,u-s) + \frac{1}{2\pi s^2}\right) \right] 
\ , 
\label{eq:F1-final-nice}
\end{align}
where $\Theta$ is the Heaviside step function and 
\begin{align}
\mathcal{A}(\tau',\tau'') 
&:= 
\partial_{\tau'}
\partial_{\tau''}
\mathcal{W}(\tau',\tau'')
\ . 
\label{eq:A-distr}
\end{align} 
The last term in \eqref{eq:F1-final-nice} does not require 
a distributional treatment at $s=0$ because 
of the subtraction ${(2\pi s^2)}^{-1}$. 
The price for this subtraction is the emergence of the first two 
terms in~\eqref{eq:F1-final-nice}, 
neither of which depends on the quantum 
state of the field or on the detector's motion. 

For $\mathcal{F}^{(p)}$ with $p>1$, expressions similar to 
\eqref{eq:F0-final-nice}
and 
\eqref{eq:F1-final-nice}
can be 
obtained by the techniques of~\cite{hodgkinson-louko}. 
We shall consider only $\mathcal{F}^{(0)}$ and $\mathcal{F}^{(1)}$.

\section{Inertial detector in 
$1+1$ Minkowski\label{sec:minkowski}} 

Let $M$ denote two-dimensional Minkowski spacetime, with the metric
$ds^2 = -dt^2 + dx^2$ in standard global Minkowski coordinates
$(t,x)$. We may alternatively use the global null coordinates $u :=
t-x$ and $v := t+x$, in which $ds^2 = - du \, dv$.  

We consider a massless scalar field. The Wightman function in the
usual Minkowski vacuum $|0_M\rangle$ is
\begin{align}
\langle 0_M |
\phi(\mathsf{x}) \phi(\mathsf{x}') | 0_M \rangle
& =
- {(4\pi)}^{-1}
\ln 
\left[
m_0 ( \epsilon + i \Delta u) 
\right] 
- {(4\pi)}^{-1}
\ln 
\left[
m_0 ( \epsilon + i \Delta v) 
\right] 
\ , 
\label{eq:mzero-Wightman-Mink}
\end{align}
where $\Delta u = u - u'$, $\Delta v = v - v'$, $m_0$ is a positive
constant of dimension inverse length, the logarithms have their
principal branch, and the distributional sense is that of $\epsilon
\to 0_+$. 
Because the field is massless, the right-moving and left-moving parts
decouple: the $\Delta u$-dependent term in
\eqref{eq:mzero-Wightman-Mink} comes from the right-movers and the
$\Delta v$-dependent term comes from the left-movers. 

The constant $m_0$ can be understood as an infrared frequency  
cutoff, and its presence renders the 
Wightman function ambiguous by an additive real-valued constant. 
From \eqref{eq:respfunc-def} it is seen that $\mathcal{F}^{(0)}$ 
in $|0_M\rangle$
depends on $m_0$ via the additive term 
\begin{align}
- \frac{\ln(m_0)}{\pi}
\int^{\infty}_{0} 
ds \, \cos(\omega s) 
\int_{-\infty}^{\infty} d u \, 
\chi(u) \chi(u-s) 
\ , 
\label{eq:F0-ir-ambig}
\end{align}
and the response of the $p=0$ detector is hence infrared ambiguous. 
The response of each of the $p>0$ detectors is however infrared unambiguous 
since the additive constant in the Wightman function 
drops out on taking the derivatives in~\eqref{eq:respfunc-def}. 

For an inertial trajectory, we have
$\mathcal{W}(\tau', \tau'') = - {(2\pi)}^{-1} \ln \bigl[ m_0 \bigl(
\epsilon + i (\tau'-\tau'') \bigr) \bigr]$, and 
\eqref{eq:F0-final-nice} and \eqref{eq:F1-final-nice} give
\begin{subequations}
\label{eq:Fboth-0M-inert}
\begin{align}
{}^{\text{in}}\mathcal{F}^{(0)}_{|0_M\rangle}(\omega)
&=
- 
\int^{\infty}_{0} 
ds 
\left[ \tfrac12 \sin(\omega s) + \pi^{-1} \cos(\omega s) \ln(m_0 s) \right]
\int_{-\infty}^{\infty} d u \, 
\chi(u)\chi(u-s)  
\ , 
\label{eq:F0-0M-inert}
\\[1ex]
{}^{\text{in}}\mathcal{F}^{(1)}_{|0_M\rangle}(\omega)
&=
-\omega \Theta(-\omega) 
\int_{-\infty}^{\infty} d u \, {[\chi(u)]}^2
\notag
\\[1ex]
&\hspace{3ex}
+ 
\frac{1}{\pi}
\int^{\infty}_{0} 
ds \, 
\frac{\cos(\omega s)}{s^2} 
\int_{-\infty}^{\infty} d u \, 
\chi(u) [\chi(u) - \chi(u-s)] 
\ , 
\label{eq:F1-0M-inert}
\end{align}
\end{subequations}
where the left superscript ${}^{\text{in}}$ indicates 
that the trajectory is inertial. 
At a large energy gap, 
$|\omega|\to\infty$, we show in 
Appendix \ref{app:asymptotics} that 
\begin{subequations}
\label{eq:Fboth-0M-inert-large}
\begin{align}
{}^{\text{in}}\mathcal{F}^{(0)}_{|0_M\rangle}(\omega)
&= 
-\frac{\Theta(-\omega)}{\omega} 
\left[
\int_{-\infty}^{\infty} d u \, 
{[\chi(u)]}^2 
\ + 
\frac{1}{\omega^2}
\int_{-\infty}^{\infty} d u
\left[\chi'(u)\right]^2 
\right.
\notag
\\[1ex]
&
\hspace{15ex}
\left.
+ \cdots + 
\frac{1}{\omega^{2k}}
\int_{-\infty}^{\infty} d u 
\left[\chi^{(k)}(u)\right]^2 
\right]
\ + O\!\left(\frac{1}{\omega^{2k+3}}\right)
\ , 
\label{eq:F0-0M-inert-large}
\\[1ex]
{}^{\text{in}}\mathcal{F}^{(1)}_{|0_M\rangle}(\omega)
&=
-\omega \Theta(-\omega) 
\int_{-\infty}^{\infty} d u \, {[\chi(u)]}^2
\ + O\!\left(\frac{1}{\omega^{2k}}\right)
\ , 
\label{eq:F1-0M-inert-large}
\end{align}
\end{subequations}
for all positive integers~$k$. 
The infrared ambiguity of 
$\mathcal{F}^{(0)}_{|0_M\rangle}$ does not 
show up in the large $|\omega|$ form 
\eqref{eq:F0-0M-inert-large} because the ambiguous contribution 
\eqref{eq:F0-ir-ambig} falls off faster 
than any inverse power of~$\omega$. 

We are also interested in the adiabatic limit of slow switching and
long detection. We implement this by writing $\chi(\tau) = g(\alpha
\tau)$ where $\alpha$ is a positive parameter, $g$ is a fixed
switching function, and the limit of interest is
$\alpha\to0_+$. Changing integration variables by $u = v/\alpha$ and
$s = r/\alpha$, comparing with~\eqref{eq:Fboth-0M-inert-large}, and
assuming $\omega\ne0$, we see that
\begin{subequations}
\label{eq:Fboth-0M-inert-adiab-fin}
\begin{align}
{}^{\text{in}}\mathcal{F}^{(0)}_{|0_M\rangle}(\omega)
&= 
-\frac{\Theta(-\omega)}{\omega} 
\left[
\alpha^{-1} \int_{-\infty}^{\infty} d v \, 
{[g(v)]}^2 
\ + 
\frac{\alpha}{\omega^2}
\int_{-\infty}^{\infty} d v
\left[g'(v)\right]^2 
\right.
\notag
\\[1ex]
&
\hspace{15ex}
\left.
+ \cdots + 
\frac{\alpha^{2k-1}}{\omega^{2k}}
\int_{-\infty}^{\infty} d v 
\left[g^{(k)}(v)\right]^2 
\right]
\ + O \bigl(\alpha^{2k+1}\bigr)
\ , 
\label{eq:F0-0M-inert-adiab-fin}
\\[1ex]
{}^{\text{in}}\mathcal{F}^{(1)}_{|0_M\rangle}(\omega)
&=
-\omega \Theta(-\omega) 
\, \alpha^{-1} 
\int_{-\infty}^{\infty} d v \, {[g(v)]}^2
\ + O\bigl(\alpha^{2k}\bigr)
\ , 
\label{eq:F1-0M-inert-adiab-fin}
\end{align}
\end{subequations}
for all positive integers~$k$. 
The probability of an excitation hence vanishes in the 
adiabatic limit, 
while the probablity of a de-exitation is proportional 
to~$\alpha^{-1}$, that is, proportional to the total detection
time.

\section{A Rindler firewall state 
$\hat\rho_{\text{\scriptsize\rm FW}}$\label{sec:firewallstate-def}}

We continue to consider a massless scalar field $\phi$ 
on two-dimensional Minkowski spacetime~$M$, 
in the notation of Section~\ref{sec:minkowski}. 
In this section we construct a mixed state $\hat\rho_{\text{FW}}$ 
in which correlations that are present in $|0_M\rangle$ 
have been severed across the Rindler horizon. 
We discuss the sense in which $\hat\rho_{\text{FW}}$ 
models the stationary aspects of a similar severing 
that has been argued in \cite{Almheiri:2012rt} to ensue 
dynamically in an evaporating 
black hole spacetime. 

\subsection{Definition of $\hat\rho_{\text{\scriptsize\rm FW}}$}

\begin{table}[p]
\begin{center}
  \begin{tabular}{ l | l | l }
Quadrant & Range in $(t,x)$ & Range in $(u,v)$\\
\hline
{\bfseries F}: \ future
& $t > |x|$ 
& $u>0$, $v>0$
\\ 
{\bfseries P}: \ past
& $t <  - |x|$ 
& $u<0$, $v<0$
\\ 
{\bfseries R}: \ right
& $x > |t|$ 
& $u<0$, $v>0$
\\ 
{\bfseries L}: \ \hspace{.5ex}left
& $x < -|t|$ 
& $u>0$, $v<0$
\end{tabular}
\end{center}
\caption{The four open quadrants of 
two-dimensional Minkowski spacetime.\label{table:quadrants}}
\end{table}

\begin{figure}[p]
\centering
\includegraphics[width=0.7\textwidth]{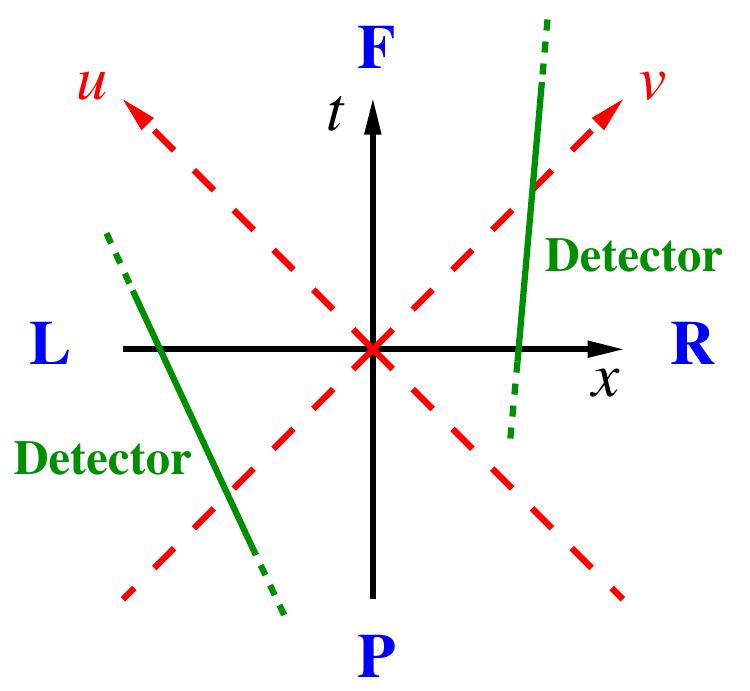}
\caption{$(1+1)$-dimensional Minkowski spacetime. 
The dashed (red) lines show the Rindler horizon $t^2-x^2=0$, 
which separates the 
four quadrants 
{\bfseries F}, 
{\bfseries P}, 
{\bfseries R} 
and 
{\bfseries L}
as summarised in Table~\ref{table:quadrants}. 
Also shown are the worldlines (green) of two inertial detectors, 
each of which operates for a finite interval of time and 
crosses during that interval exactly 
one branch of the Rindler horizon.}
\label{fig:minkowski}
\end{figure}

Recall that the Rindler horizon in $M$ is at 
$t^2-x^2=0$, or in terms of the null
coordinates, at $uv=0$.  We denote the future, past, right and left
open quadrants separated by the Rindler horizon by respectively
{\bfseries F}, {\bfseries P}, {\bfseries R} and~{\bfseries L}, as
summarised in Table \ref{table:quadrants} and shown in
Figure~\ref{fig:minkowski}.

Recall also that the restriction of $|0_M\rangle$ to {\bfseries R} is
a mixed state whose density matrix $\rho_{\text{{\bfseries R}}}$ is
thermal in temperature ${(2\pi)}^{-1}$ with respect to the boost
Killing vector $\xi := x\partial_t + t\partial_x 
= - u\partial_u + v\partial_v$, which is timelike
and future-pointing in {\bfseries R}
\cite{unruh,birrell-davies,wald-smallbook}.  Similarly, the
restriction of $|0_M\rangle$ to {\bfseries L} is a mixed state whose
density matrix $\rho_{\text{{\bfseries L}}}$ is thermal in temperature
${(2\pi)}^{-1}$ with respect to the boost Killing vector $-\xi$, which
is timelike and future-pointing in~{\bfseries L}\null.

Now, consider on $\text{\bfseries R} \cup \text{\bfseries L}$ the
mixed state whose density matrix is $\rho_{\text{FW}} :=
\rho_{\text{{\bfseries R}}} \otimes \rho_{\text{{\bfseries L}}}$.  For
any observable whose support is contained in~{\bfseries R}, the
expectation value in $\rho_{\text{FW}}$ is identical to the
expectation value in~$|0_M\rangle$, and similarly for any observable
whose support is contained in~{\bfseries L}\null.  However,
$\rho_{\text{FW}}$ contains no correlations between {\bfseries R}
and~{\bfseries L}: all the correlations between {\bfseries R} and
{\bfseries L} that are present in $|0_M\rangle$
\cite{unruh,birrell-davies,wald-smallbook,reznik1} have been severed
in~$\rho_{\text{FW}}$.

We wish to extend $\rho_{\text{FW}}$ beyond $\text{\bfseries R} \cup
\text{\bfseries L}$.  There exists a unique extension to
$\text{\bfseries F} \cup \text{\bfseries P} \cup \text{\bfseries R}
\cup \text{\bfseries L}$: because the field is massless, the
left-moving part of the field propagates into {\bfseries F} only from
{\bfseries R} and into {\bfseries P} only from {\bfseries L}, while
the right-moving part of the field propagates into {\bfseries F} only
from {\bfseries L} and into {\bfseries P} 
only from~{\bfseries R}\null. 
We denote this extension by $\tilde\rho_{\text{FW}}$.  The Wightman
function in $\tilde\rho_{\text{FW}}$, given by
$\Tr\bigl(\phi(\mathsf{x}) \phi(\mathsf{x}') \tilde\rho_{\text{FW}}
\bigr)$, is equal to $\langle 0_M | \phi(\mathsf{x}) \phi(\mathsf{x}')
| 0_M \rangle$ when $\mathsf{x}$ and $\mathsf{x}'$ are in the same
quadrant, but not when $\mathsf{x}$ and $\mathsf{x}'$ are in distinct
quadrants, as collected in Table~\ref{table:FW-Wightman}.

\begin{table}[t]
\begin{center}
\begin{tabular}{ l | l }
Quadrant pairs & 
$\vphantom{{\displaystyle\frac{A}{A}}}
\Tr\bigl(\phi(\mathsf{x}) \phi(\mathsf{x}') \tilde\rho_{\text{FW}} \bigr)$\\
\hline
$\vphantom{{\displaystyle\frac{\text{\huge A}}{\text{\huge A}}}}
\genfrac{}{}{0pt}{0}{\text{{\bfseries P} and {\bfseries R}}}{\text{{\bfseries L} and {\bfseries F}}}$ 
& $- {(4\pi)}^{-1}
\ln 
\left[
m_0 ( \epsilon + i \Delta u) 
\right] 
$ 
\\ 
\hline
$\vphantom{{\displaystyle\frac{\text{\huge A}}{\text{\huge A}}}}
\genfrac{}{}{0pt}{0}{\text{{\bfseries P} and {\bfseries L}}}{\text{{\bfseries R} and {\bfseries F}}}$ 
& $- {(4\pi)}^{-1}
\ln 
\left[
m_0 ( \epsilon + i \Delta v) 
\right] 
$ 
\\ 
\hline
$\vphantom{{\displaystyle\frac{\text{\huge A}}{\text{\huge A}}}}
\genfrac{}{}{0pt}{0}{\text{{\bfseries R} and {\bfseries L}}}{\text{{\bfseries P} and {\bfseries F}}}$ 
& $0$ 
\\ 
\hline
\end{tabular}
\end{center}
\caption{The table shows 
$\Tr\bigl(\phi(\mathsf{x}) \phi(\mathsf{x}') \tilde\rho_{\text{FW}} \bigr)$ 
when $\mathsf{x}$ and $\mathsf{x}'$ 
are in distinct quadrants of 
$\text{\bfseries F} \cup \text{\bfseries P} 
\cup \text{\bfseries R} \cup \text{\bfseries L}$\null. 
In the pairs 
$(\text{\bfseries P}, \text{\bfseries R})$ 
and 
$(\text{\bfseries L}, \text{\bfseries F})$, 
the two quadrants 
are causally correlated for right-movers 
and $\Tr\bigl(\phi(\mathsf{x}) \phi(\mathsf{x}') \tilde\rho_{\text{FW}} \bigr)$ 
contains only the right-mover contribution to  
$\langle 0_M | \phi(\mathsf{x}) \phi(\mathsf{x}') | 0_M \rangle$. 
In the pairs  
$(\text{\bfseries P}, \text{\bfseries L})$ 
and 
$(\text{\bfseries R}, \text{\bfseries F})$, 
the two quadrants 
are causally correlated for left-movers 
and $\Tr\bigl(\phi(\mathsf{x}) \phi(\mathsf{x}') \tilde\rho_{\text{FW}} \bigr)$ 
contains only the left-mover contribution to 
$\langle 0_M | \phi(\mathsf{x}) \phi(\mathsf{x}') | 0_M \rangle$. 
In the pairs 
$(\text{\bfseries R}, \text{\bfseries L})$ 
and 
$(\text{\bfseries P}, \text{\bfseries F})$, the two quadrants have 
no causal correlation and 
$\Tr\bigl(\phi(\mathsf{x}) \phi(\mathsf{x}') \tilde\rho_{\text{FW}} \bigr)$
vanishes. 
When $\mathsf{x}$ and $\mathsf{x}'$ are in the same quadrant, 
$\Tr\bigl(\phi(\mathsf{x}) \phi(\mathsf{x}') \tilde\rho_{\text{FW}} \bigr)
= 
\langle 0_M |
\phi(\mathsf{x}) \phi(\mathsf{x}') 
| 0_M \rangle$.\label{table:FW-Wightman}}
\end{table}

Extending $\Tr\bigl(\phi(\mathsf{x})
\phi(\mathsf{x}') \tilde\rho_{\text{FW}} \bigr)$ from 
$\text{\bfseries F} \cup
\text{\bfseries P} \cup \text{\bfseries R} \cup \text{\bfseries L}$
to all of Minkowski requires additional input on the Rindler horizon. 
We adopt the extension that is minimal in the sense that it has no 
distributional support at the Rindler horizon. 
This extension is unique: 
we denote it by 
$\Tr\bigl(\phi(\mathsf{x}) \phi(\mathsf{x}') \hat\rho_{\text{FW}}
\bigr)$, and we interpret it as the Wightman function of a mixed state
whose density matrix we denote by~$\hat\rho_{\text{FW}}$.

\subsection{Properties of $\hat\rho_{\text{\scriptsize\rm FW}}$}

By construction, $\hat\rho_{\text{FW}}$ is indistinguishable from 
$| 0_M \rangle$ for any operator whose support 
is contained in any one of the four quadrants \text{\bfseries F}, 
\text{\bfseries P}, \text{\bfseries R}, and~\text{\bfseries L}\null. 
In particular, the restriction of $\hat\rho_{\text{FW}}$ 
to any one of the quadrants is Hadamard and has a
vanishing stress-energy tensor. 

The restriction of $\hat\rho_{\text{FW}}$ 
to any one of the quadrants is also invariant under the Lorentz boosts 
generated by the Killing vector~$\xi$. 
The restrictions of $\hat\rho_{\text{FW}}$ to $\text{\bfseries R}$ and 
$\text{\bfseries L}$ are hence stationary with respect to Rindler time translations, 
and observers on the uniformly-accelerated 
world lines $x^2 - t^2 = a^{-2}$, 
where the positive constant $a$ is the acceleration, 
will experience the usual Unruh effect, in temperature 
$a/(2\pi)$ \cite{unruh,Crispino:2007eb}. 
The restrictions of $\hat\rho_{\text{FW}}$ to $\text{\bfseries R}$ and 
$\text{\bfseries L}$ are also invariant with respect to 
Minkowski time translations in a local sense, 
but not globally, since Minkowski time translations necessarily map 
$\text{\bfseries R}$ and $\text{\bfseries L}$ 
to regions that intersect the Rindler horizons. 

The Wightman function of $\hat\rho_{\text{FW}}$ 
is by construction a well-defined distribution everywhere, 
including the Rindler horizon. 
The response of a horizon-crossing detector in the state $\hat\rho_{\text{FW}}$ 
is hence well defined by~\eqref{eq:respfunc-def}. 
As the Wightman function 
is not invariant under Lorentz boosts generated by $\xi$ when the two arguments are in 
distinct quadrants of the pairs 
$(\text{\bfseries P}, \text{\bfseries R})$, 
$(\text{\bfseries P}, \text{\bfseries L})$, 
$(\text{\bfseries R}, \text{\bfseries F})$ or 
$(\text{\bfseries L}, \text{\bfseries F})$, 
we may expect Lorentz-noninvariance 
in the response of a detector that crosses 
exactly one branch of the Rindler horizon, 
and we may expect this noninvariance 
to be associated with the infrared cutoff~$m_0$: 
this is what will be found in Section~\ref{sec:resp-across}. 

The Wightman function of $\hat\rho_{\text{FW}}$ 
is not Hadamard at the Rindler horizon. 
We shall not attempt to examine in which sense 
$\hat\rho_{\text{FW}}$ may be definable on the Rindler horizon 
beyond its Wightman function, and 
in particlar we shall not attempt to define 
a stress-energy tensor for 
$\hat\rho_{\text{FW}}$ on the Rindler horizon. 
We shall return to this point in Section~\ref{sec:conc}.

\subsection{$\hat\rho_{\text{\scriptsize\rm FW}}$ as a firewall model}

$\hat\rho_{\text{FW}}$ contains by construction no correlations
between the spacetime regions {\bfseries R} and~{\bfseries L}\null. 
We may view $\hat\rho_{\text{FW}}$ as the minimal 
modification of $| 0_M \rangle$ in which the correlations 
between {\bfseries R} and {\bfseries L}
\cite{unruh,birrell-davies,wald-smallbook,reznik1} have been
fully severed. 
The severing has made $\hat\rho_{\text{FW}}$ singular on the
Rindler horizon, but with a Wightman function that is still a
well-defined distribution.

In the spacetime of an evaporating black hole, the conventional 
quantum field theory picture implies that
the field develops strong correlations between the interior
and exterior of the hole, closely similar to the correlations in $|
0_M \rangle$ across the Rindler horizon
\cite{hawking,unruh,birrell-davies,wald-smallbook}.  
It is argued in \cite{Almheiri:2012rt} 
that these correlations cannot
be maintained if the quantum evolution of the full system is assumed
unitary. It is further argued in \cite{Almheiri:2012rt} 
that breaking the correlations will
replace the horizon by a firewall, a region of high curvature,
which will destroy any observer who attempts to fall into the black
hole. Our state $\hat\rho_{\text{FW}}$ models within Minkowski
spacetime quantum field theory the severed quantum correlations across
the firewall of~\cite{Almheiri:2012rt}. A~detector crossing the
Rindler horizon from {\bfseries R} or {\bfseries L} to {\bfseries F},
in the state~$\hat\rho_{\text{FW}}$, models a
detector crossing the firewall of \cite{Almheiri:2012rt} as long as
the shrinking black hole horizon has not yet become gravitationally
singular due to any back-reaction from the stress-energy 
of the firewall quantum state. 

In summary, $\hat\rho_{\text{FW}}$ models the
stationary aspects of the black hole firewall of~\cite{Almheiri:2012rt}.  The
relevant sense of stationarity in $\hat\rho_{\text{FW}}$ is with
respect to Lorentz boosts. The Lorentz-nonivariance
of $\hat\rho_{\text{FW}}$ means that the modelling will not be fully
stationary, but this nonstationarity is associated with the infrared
cutoff $m_0$ and we will see that it will not be significant for the
conclusions.

We emphasise that we shall not attempt to model how the severing of
the quantum correlations in the firewall state of
\cite{Almheiri:2012rt} may arise through the evolution of the full
quantum system, nor shall we attempt to model how the spacetime reacts
to the singularity in the firewall state. Also, we shall not attempt
to discuss in detail the near-horizon phenomena proposed
in~\cite{braunstein-et-al,Mathur:2009hf}, 
but we shall consider in Section \ref{sec:curtain}
a generalisation of $\hat\rho_{\text{FW}}$ that has been suggested 
\cite{braunstein-private} to model the energetic curtain 
of~\cite{braunstein-et-al}. 

We refer to $\hat\rho_{\text{FW}}$ as a Rindler firewall state.

\section{Response of an inertial detector in 
$\hat\rho_{\text{\scriptsize\rm FW}}$\label{sec:resp-across}}

In this section we examine the response of the two-level
detector of Section \ref{sec:detector} when it 
crosses the Rindler horizon and the field is in the 
Rindler firewall state~$\hat\rho_{\text{FW}}$. 
We take the detector to be inertial and to cross the horizon 
exactly once during the time that it operates. 
Subsection \ref{subsec:genpointcrossing} 
considers the generic case, shown in Figure~\ref{fig:minkowski}, in which 
the horizon-crossing occurs away from the 
bifurcation point $(t,x) = (0,0)$. 
Crossing from {\bfseries R} or {\bfseries L} to {\bfseries F} 
models crossing the black hole firewall of~\cite{Almheiri:2012rt}, but 
we shall see that crossing from 
{\bfseries P} to {\bfseries R} or~{\bfseries L} yields an identical response.  
The special case of a detector that goes through 
the bifurcation point is treated in subsection~\ref{subsec:bifpointcrossing}.

\subsection{Generic horizon-crossing\label{subsec:genpointcrossing}}

In this subsection we consider an inertial detector that crosses 
exactly one branch of the horizon during the time that it 
operates, as shown in Figure~\ref{fig:minkowski}. 
We introduce the parameter $\eta$ that takes the value $1$ 
if this this branch is the left-going branch, $v=0$, and the value 
$-1$ if this branch is the right-going branch, $u=0$. 
We write the detector's velocity vector as 
$\cosh(\beta) \partial_t + \sinh(\beta) \partial_x$,
where $\beta \in \BbbR$ is the rapidity with respect to the
Lorentz-frame $(t,x)$, and we parametrise the trajectory so that
the horizon-crossing occurs at $\tau=0$. 

Let ${}^{\text{in}}\mathcal{F}^{(p)}_{|0_M\rangle}$ and 
${}^{\text{in}}\mathcal{F}^{(p)}_{\text{FW}}$ 
denote the response of the inertial detector
in the respective states $| 0_M \rangle$ and~$\hat\rho_{\text{FW}}$. 
Using \eqref{eq:respfunc-def}, and the Wightman functions given in 
\eqref{eq:mzero-Wightman-Mink} and in Table~\ref{table:FW-Wightman}, we see  
that the difference 
$\Delta\mathcal{F}^{(p)} := {}^{\text{in}}\mathcal{F}^{(p)}_{\text{FW}} 
- {}^{\text{in}}\mathcal{F}^{(p)}_{|0_M\rangle}$
is given by 
\begin{align}
\label{eq:DeltaF-def}
\Delta \mathcal{F}^{(p)}(\omega)
&=
\int^{\infty}_{-\infty}\,d\tau'\,\int^{\infty}_{-\infty}\,d\tau''\, 
e^{-i \omega(\tau'-\tau'')} \,\chi(\tau')\chi(\tau'') \, 
\partial^p_{\tau'}
\partial^p_{\tau''}
\Delta \mathcal{W}(\tau',\tau'')
\ , 
\end{align}
where 
\begin{align}
\label{eq:DeltaW-def}
\Delta \mathcal{W}(\tau',\tau'')
= 
\begin{cases}
{(4\pi)}^{-1} \ln \bigl[m_0 e^{\eta\beta} (\tau'-\tau'') \bigr] 
+ \tfrac18 i 
& \text{for $\tau'>0>\tau''$}
\ , 
\\
{(4\pi)}^{-1} \ln \bigl[m_0 e^{\eta\beta} (\tau''-\tau') \bigr] 
- \tfrac18 i 
& \text{for $\tau''>0>\tau'$}
\ , 
\\
0 
& \text{otherwise}\ . 
\end{cases}
\end{align}
$\Delta \mathcal{F}^{(p)}$ is well defined and finite for each~$p$: 
the derivatives in
\eqref{eq:DeltaF-def} are distributional 
but the integrals exist and are finite 
since $\chi$ is by assumption smooth and of compact
support. 

We now specialise to $p=0$ and $p=1$. It is shown in Appendix 
\ref{app:evaluation} that 
\begin{subequations}
\label{eq:DeltaFboth}
\begin{align}
\Delta \mathcal{F}^{(0)}(\omega)
&= 
\int_0^\infty 
\,ds 
\left[ \tfrac14 \sin(\omega s) + {(2\pi)}^{-1} \cos(\omega s) 
\ln(m_0 e^{\eta\beta}s) \right]
\, \int_0^s du \, 
\chi(u) \chi(u-s) 
\ , 
\label{eq:DeltaF0}
\\[1ex]
\Delta \mathcal{F}^{(1)}(\omega)
& =
\frac{{[\chi(0)]}^2}{2\pi}
\ln \bigl( |\omega|e^{\gamma-1} e^{-\eta\beta}/m_0\bigr)
\notag
\\[1ex]
& 
\hspace*{2ex}
+ 
\frac{1}{2\pi}
\int_0^\infty 
\,ds \cos(\omega s) 
\, \Biggl\{
\frac{\chi(0) \bigl[
\chi(0) - \chi(s) - \chi(-s)
\bigr]}{s} 
\notag
\\[1ex]
& 
\hspace*{26ex}
+ 
\frac{1}{s^2}
\int_0^s du \, 
\chi(u) \chi(u-s) 
\Biggr\}
\ \ \ \text{for $\omega\ne0$}\ , 
\label{eq:DeltaF1-nonzero}
\\[1ex]
\Delta \mathcal{F}^{(1)}(0)
& =
\frac{\chi(0)}{4\pi}
\int_0^\infty 
\,ds \ln( e m_0 e^{\eta\beta}s) 
\bigl[
\chi'(s) - \chi'(-s) \bigr]
\notag
\\[1ex]
& 
\hspace*{2ex}
+ 
\frac{1}{2\pi}
\int_0^\infty 
\,ds 
\left\{
- \frac{\chi(0) \bigl[\chi(s) + \chi(-s)
\bigr]}{2s} 
+ 
\frac{1}{s^2}
\int_0^s du \, 
\chi(u) \chi(u-s) 
\right\}
\ ,
\label{eq:DeltaF1-zero}
\end{align}
\end{subequations}
where $\gamma$ is Euler's constant. 

Several observations are in order. 
For properties that hold for both of 
$\Delta \mathcal{F}^{(0)}$ and $\Delta \mathcal{F}^{(1)}$, 
we refer to the two by~$\Delta \mathcal{F}$. 


First, $\Delta \mathcal{F}^{(1)}$ is even in~$\omega$: 
the firewall has identical effects on probabilities 
of excitation and de-excitation 
of the derivative-coupling detector. 
By contrast, $\Delta \mathcal{F}^{(0)}$ 
has no fixed parity
in~$\omega$, but we shall see below
that the dominant contribution to $\Delta \mathcal{F}^{(0)}$
at large $|\omega|$ is even in~$\omega$.\footnote{
This paragraph differs from the JHEP version, 
correcting the parity description of~$\Delta \mathcal{F}^{(0)}$.}

Second, $\Delta \mathcal{F}$ is invariant under $\chi(\tau) \to
\chi(-\tau)$: the firewall effect is invariant under a future-past
reflection about the horizon-crossing moment.

Third, $\Delta \mathcal{F}$ depends on the infrared cutoff~$m_0$. 
It also depends on the trajectory's rapidity parameter $\beta$ 
and is hence not Lorentz invariant. We shall shortly see that 
these effects are subdominant in the limit of a large energy 
gap and in the limit of adiabatic switching, 
but we may note here that the Lorentz noninvariance is directly 
connected to the cutoff: 
the term that depends on $m_0$ and $\beta$ is 
\begin{subequations}
\begin{align}
p=0: & \hspace{4ex}
\frac{\ln(m_0 e^{\eta\beta})}{2\pi}
\int_0^\infty 
\,ds \cos(\omega s)  
\, \int_0^s du \, 
\chi(u) \chi(u-s)
\ , 
\label{eq:F0-diff-m0term}
\\[1ex]
p=1: & \hspace{4ex}
-{(2\pi)}^{-1} {[\chi(0)]}^2
\ln(m_0 e^{\eta\beta})
\ , 
\label{eq:F1-diff-m0term}
\end{align}
\end{subequations}
which shows that 
increasing (respectively decreasing) the detector's velocity towards
the horizon has the effect of increasing (decreasing) the effective
infrared cutoff $m_0 e^{\eta\beta}$ by precisely the appropriate
Doppler shift factor. Note also that for $p=0$ 
the ambiguous term \eqref{eq:F0-diff-m0term} 
comes from a finite neighbourhood of the horizon-crossing moment, 
while for $p=1$ 
the ambiguous term \eqref{eq:F1-diff-m0term} 
comes strictly from the horizon-crossing moment and vanishes 
iff $\chi(0)=0$. 

Fourth, $\Delta \mathcal{F}$ is nonvanishing whenever $\chi$ has support both 
before and after the horizon-crossing, 
regardless whether the detector is in operation 
at the horizon-crossing moment. 

Fifth, we show in Appendix \ref{app:asymptotics} 
that $\Delta \mathcal{F}$ has the large 
$|\omega|$ form 
\begin{subequations}
\label{eq:deltaFboth-largeomega}
\begin{align}
\Delta \mathcal{F}^{(0)}(\omega)
&=
\frac{1}{2\pi}\ln \bigl( |\omega|e^{\gamma-1} e^{-\eta\beta}/m_0\bigr)
\left(\frac{{[\chi(0)]}^2}{\omega^2} + 
\frac{{[\chi'(0)]}^2 -  2 \chi(0) \chi''(0)}{\omega^4} 
+ O\bigl(\omega^{-6}\bigr)
\right)
\notag
\\[1ex]
& \hspace{3ex}
+ \frac{2 \chi(0) \chi''(0) - {[\chi'(0)]}^2}{6 \pi \omega^4} 
+ O\bigl(\omega^{-6}\bigr)
\ , 
\label{eq:deltaF0-largeomega}
\\[1ex]
\Delta \mathcal{F}^{(1)}(\omega)
&= \frac{{[\chi(0)]}^2}{2\pi}
\ln \bigl( |\omega|e^{\gamma-1} e^{-\eta\beta}/m_0\bigr)
+ \frac{ 4 \chi(0) \chi''(0) + {[\chi'(0)]}^2}{12 \pi \omega^2} 
+ O\bigl(\omega^{-4}\bigr)
\ . 
\label{eq:deltaF1-largeomega}
\end{align}
\end{subequations}
In the special case $\chi(0)=0$, all the terms shown in 
\eqref{eq:deltaFboth-largeomega} 
vanish and the first potentially nonvanishing terms are 
\begin{subequations}
\label{eq:deltaFboth-deg-largeomega}
\begin{align}
\Delta \mathcal{F}^{(0)}(\omega)
&=
\frac{1}{2\pi}\ln \bigl( |\omega|e^{\gamma-1} e^{-\eta\beta}/m_0\bigr)
\left(\frac{{[\chi''(0)]}^2}{\omega^6} + O\bigl(\omega^{-8}\bigr)
\right)
- \frac{8{[\chi''(0)]}^2}{30 \pi \omega^6}
+ O\bigl(\omega^{-8}\bigr)
\ , 
\label{eq:deltaF0-deg-largeomega}
\\
\Delta \mathcal{F}^{(1)}(\omega)
&= 
\frac{{[\chi''(0)]}^2}{40 \pi \omega^4} + O\bigl(\omega^{-6}\bigr)
\ . 
\label{eq:deltaF1-deg-largeomega}
\end{align}
\end{subequations}
The dominant effect at large $|\omega|$ comes hence from the
horizon-crossing moment.  If $\chi$ and all its derivatives vanish at
the horizon-crossing, $\Delta \mathcal{F}$ 
vanishes at $|\omega|\to\infty$ faster than any
inverse power of~$\omega$.

Sixth, to analyse the adiabatic limit, we write $\chi(\tau) = g(\alpha
\tau)$ where $\alpha$ is a positive parameter, $g$ is a fixed
switching function, and we are interested in the limit
$\alpha\to0_+$. Changing in \eqref{eq:DeltaFboth}
integration variables by $u = v/\alpha$ and
$s = r/\alpha$, and assuming $\omega\ne0$, we see that the asymptotic
formulas are obtained from \eqref{eq:deltaFboth-largeomega} and
\eqref{eq:deltaFboth-deg-largeomega} by multiplying $\Delta
\mathcal{F}^{(0)}$ by $\alpha^{-2}$ and making the replacements $\chi
\to g$, $\omega \to \omega/\alpha$ and $m_0 \to m_0/\alpha$. The
dominant effect in $\Delta \mathcal{F}$ in the adiabatic limit hence comes from the
horizon-crossing moment, and if the detector operates at this moment,
the leading term in $\Delta \mathcal{F}$ 
is independent of $\alpha$ and equal to the leading 
term shown in~\eqref{eq:deltaFboth-largeomega}. 
When the Minkowski vacuum contribution 
\eqref{eq:Fboth-0M-inert-adiab-fin} to the response is included, 
we see that if the detector operates at the horizon-crossing moment, 
the firewall 
gives the leading adiabatic contribution to the 
excitation probability and the next-to-leading 
adiabatic contribution to the de-excitation probability.

\subsection{Horizon-crossing at the 
bifurcation point\label{subsec:bifpointcrossing}}

In the special case in which the detector 
crosses the horizon at the bifurcation point, 
$\Delta \mathcal{F}$ is given by summing over 
the two values of $\eta$ in~\eqref{eq:DeltaFboth}. 
$\Delta \mathcal{F}$ is hence obtained from \eqref{eq:DeltaFboth}
by setting $\eta=0$ and including an 
overall multiplicative factor~$2$. 
The only qualitatively new property is that 
$\Delta \mathcal{F}$ 
is now independent of $\beta$ and hence Lorentz invariant.

\section{Rindler energetic curtain\label{sec:curtain}}

In this section we consider a generalisation of $\hat\rho_{\text{FW}}$
whose restriction to {\bfseries L} is thermal with respect to the
future-pointing Killing vector $-\xi$ in the (dimensionless)
temperature $T>0$, and an inertial detector crossing the Rindler horizon 
from {\bfseries R} to~{\bfseries F}\null. 
It has been suggested \cite{braunstein-private} 
that at $T\gg{(2\pi)}^{-1}$ this system models a detector 
crossing the energetic curtain of
\cite{braunstein-et-al} in a black hole spacetime.

\subsection{The state}

Let $\Mhat$ denote an auxiliary $(1+1)$-dimensional Minkowski
spacetime, with the metric 
$\widehat{ds^2} = - d\hat u \, d\hat v$ in the dimensionless null coordinates 
$(\hat u,\hat v)$. 
For a massless scalar field on~$\Mhat$, the
Wightman function in a thermal state of temperature $T>0$
with respect to the normalised time translation Killing vector
$\hat\xi := \partial_{\hat u} + \partial_{\hat v}$ 
reads \cite{Juarez-Aubry:2014jba}
\begin{align}
{\hat G}_T\bigl(({\hat u}', {\hat v}'), ({\hat u}'', {\hat v}'') \bigr) 
= 
-{(4 \pi)}^{-1}
\ln \bigl\{
- \sinh[\pi T (\Delta\hat u- i\epsilon)]
\sinh[\pi T (\Delta\hat v - i \epsilon)]
\bigr\}
\ , 
\label{eq:Mhat-GhatT-thermal}
\end{align}
where $\Delta\hat u = {\hat u}' - {\hat u}''$, $\Delta\hat v = {\hat
  v}' - {\hat v}''$, the logarithm has its principal branch
and the distributional sense is that of 
$\epsilon \to0_+$. 
Note that the temperature parameter $T$ 
is dimensionless since $\hat u$ and $\hat v$ are dimensionless. 

We map $\Mhat$ conformally to the region {\bfseries L} in the
$(1+1)$-dimensional Minkowski spacetime $M$ of  
Section~\ref{sec:minkowski}, 
by $u = m_0^{-1} e^{\hat u}$ and $v = - m_0^{-1} e^{-\hat v}$,
so that $ds^2 = - du \, dv = (-uv) \, \widehat{ds^2}$.  The push-forward
of $\hat\xi$ to {\bfseries L} 
is the future-pointing boost Killing vector $u\partial_u
- v\partial_v = -\xi$, and the push-forward of ${\hat G}_T$ is
$G_T^{\text{\bfseries L}} (\mathsf{x}, \mathsf{x}') + \frac14 T
\ln(m_0^4 u' u'' v' v'')$, where
\begin{align}
G_T^{\text{\bfseries L}} (\mathsf{x}, \mathsf{x}')
= & \, 
- {(4\pi)}^{-1}
\ln \! 
\left\{ \epsilon 
+ i \bigl[(m_0 u')^{2\pi T} - (m_0 u'')^{2\pi T}\bigr] \right\}
\notag
\\[1ex]
& \, 
- {(4\pi)}^{-1}
\ln \! 
\left\{ \epsilon 
+ i \bigl[(-m_0 v'')^{2\pi T} - (-m_0 v')^{2\pi T}\bigr] \right\}
\ . 
\end{align}
As the term $\frac14 T \ln(m_0^4 u' u'' v' v'')$ is regular in
{\bfseries L} and satisfies the field equation there, 
we may drop this term and define
in {\bfseries L} a quantum state whose Wightman function equals
$G_T^{\text{\bfseries L}}$. We denote the density matrix of this state
by $\rho_{\text{{\bfseries L}},T}$. Note that 
$\rho_{\text{{\bfseries L}},{(2\pi)}^{-1}} = \rho_{\text{\bfseries L}}$. 

Now, the right-mover part of $G_T^{\text{\bfseries L}}$ continues
without singularities from {\bfseries L} to~{\bfseries F}, and the
left-mover part continues without singularities from {\bfseries L}
to~{\bfseries P}\null. We may hence define on $M$ a state by starting from
$\rho_{\text{{\bfseries R}}} \otimes \rho_{\text{{\bfseries L},T}}$ on
$\text{\bfseries R} \cup \text{\bfseries L}$ and extending to all of
$M$ by causal propagation as in Section~\ref{sec:firewallstate-def}.
We denote the density matrix of this state by 
$\hat\rho_{\text{EC},T}$. By construction, 
$\hat\rho_{\text{EC},{(2\pi)}^{-1}} = \hat\rho_{\text{FW}}$. 
We regard $\hat\rho_{\text{EC},T}$ as modelling the energetic curtain of 
\cite{braunstein-et-al} when $T\gg{(2\pi)}^{-1}$~\cite{braunstein-private}.

\subsection{Detector}

We consider the response of an inertial detector that crosses the
Rindler horizon from {\bfseries R} to {\bfseries F}, with the field
in the state~$\hat\rho_{\text{EC},T}$. 
The response differs from that in the state $\hat\rho_{\text{FW}}$
by the additional term 
\begin{align}
\label{eq:DeltaECF-def}
\Delta_{\text{EC}}\mathcal{F}^{(p)}(\omega)
&=
\int^{\infty}_{-\infty}\,d\tau'\,\int^{\infty}_{-\infty}\,d\tau''\, 
e^{-i \omega(\tau'-\tau'')} \,\chi(\tau')\chi(\tau'') \, 
\partial^p_{\tau'}
\partial^p_{\tau''}
\Delta_{\text{EC}} \mathcal{W}(\tau',\tau'')
\ , 
\end{align}
where 
\begin{align}
\label{eq:DeltaECW-def}
\Delta_{\text{EC}} \mathcal{W}(\tau',\tau'')
= 
\begin{cases}
{\displaystyle \frac{1}{4\pi} \ln \! 
\left[
\frac{\mtilde  (\tau'-\tau'')}{{(\mtilde\tau')}^{2\pi T} - {(\mtilde\tau'')}^{2\pi T}}
\vphantom{\frac{\frac{A}{A}}{\frac{A}{A}}}
\right]}
& \text{for $\tau'>\tau''>0$ or $\tau''>\tau'>0$}\ , 
\\[2ex]
0 
& \text{otherwise}\ , 
\end{cases}
\end{align}
and $\mtilde:= m_0 e^{-\beta}$. 
$\Delta_{\text{EC}}\mathcal{F}^{(p)}$ is clearly finite for all $T$ and~$p$. 

We are interested in the limit of large~$T$. 
Proceeding as in Section~\ref{sec:resp-across}, 
and using the techniques of Appendix~\ref{app:evaluation}, 
we find 
that the asymptotic large $T$ forms of 
$\Delta_{\text{EC}}\mathcal{F}^{(0)}$
and 
$\Delta_{\text{EC}}\mathcal{F}^{(1)}$ 
are 
\begin{subequations}
\begin{align}
\Delta_{\text{EC}}\mathcal{F}^{(0)}(\omega)
&= 
- T \int_0^\infty ds \cos(\omega s) 
\int_s^\infty du \ln(\mtilde u) \, \chi(u) \chi(u-s)
\ + O(T^0) 
\ , 
\label{eq:DeltaECW0}
\\[1ex]
\Delta_{\text{EC}}\mathcal{F}^{(1)}(\omega)
&= 
T \left[ 
{[\chi(0)]}^2 \ln \! \left(\frac{|\omega| e^\gamma}{\mtilde}\right)
+ 
\chi(0) \int_0^\infty ds \cos(\omega s) \, 
\frac{\chi(0) - \chi(s)}{s}
\right.
\notag
\\[1ex]
& \hspace{7ex}
\left. 
- \int_0^\infty du \ln(\mtilde u) 
\, \chi'(u) \chi(u)
\right]
\ + O(T^0) 
\ \ \ \ \ \text{for $\omega\ne0$}\ , 
\label{eq:DeltaECW1-gen-leading}
\\[1ex]
\Delta_{\text{EC}}\mathcal{F}^{(1)}(0)
&= 
T 
\int_0^\infty du \ln(\mtilde u) 
\, \chi'(u) [\chi(0) - \chi(u)]
\ + O(T^0) 
\ . 
\label{eq:DeltaECW1-omegazero-leading}
\end{align}
\end{subequations}
The leading behaviour is hence linear in~$T$. 
When $|\omega|$ is large, 
we may use the techniques of Appendix \ref{app:asymptotics} 
to show that 
\begin{subequations}
\begin{align}
\Delta_{\text{EC}}\mathcal{F}^{(0)}(\omega)
&= 
T \left[ \frac{{[\chi(0)]}^2}{\omega^2}
\ln \! \left(\frac{|\omega|e^\gamma}{\mtilde}\right)
- \frac{1}{\omega^2}
\int_0^\infty du \ln(\mtilde u) 
\, \chi'(u) \chi(u)
\ + O \! \left(\frac{\ln(|\omega|)}{\omega^4}\right)
\right] 
\notag
\\[1ex]
& \hspace{3ex}
+ O(T^0) 
\ ,  
\label{eq:DeltaECW0-as}
\\[1ex]
\Delta_{\text{EC}}\mathcal{F}^{(1)}(\omega)
&= 
T \left[ 
{[\chi(0)]}^2 \ln \! \left(\frac{|\omega| e^\gamma}{\mtilde}\right)
- \int_0^\infty du \ln(\mtilde u) 
\, \chi'(u) \chi(u)
\ + 
\chi(0) \, O\bigl(\omega^{-2}\bigr)
\right]
\notag
\\[1ex]
& \hspace{3ex}
+ O(T^0) 
\ . 
\label{eq:DeltaECW1-as}
\end{align}
\end{subequations}
If $\chi(0)=0$, the leading $\omega$-dependence at large 
$|\omega|$ drops out 
from the $T$-term in \eqref{eq:DeltaECW0-as}, 
and the $T$-term in \eqref{eq:DeltaECW1-as} 
becomes independent of~$\omega$. 

We conclude that the response
can be made arbitrarily large by increasing~$T$, 
and the part of this response that is 
dominant at large $|\omega|$ comes from the 
horizon-crossing moment.

\section{Summary and concluding remarks\label{sec:conc}}

We have shown that a two-level UDW detector in $(1+1)$-dimensional Minkowski
spacetime, 
coupled linearly to a massless scalar field or its proper time derivative, 
has a finite response on crossing inertially the Rindler horizon in a
firewall-type quantum state in which the Minkowski vacuum correlations
between the right and left Rindler wedges have been fully severed. In
the limit of a large detector energy gap~$\omega$, the leading contribution to
the difference from the Minkowski vacuum response 
is proportional to ${[\chi(0)]}^2\omega^{-2}\ln(|\omega|)$ 
for the non-derivative detector and to ${[\chi(0)]}^2\ln(|\omega|)$ 
for the derivative-coupling detector, where 
$\chi(0)$ is the coupling strength at the horizon-crossing moment. 
The same leading contributions arise also in the 
limit of adiabatic switching. If the
detector operates both before and after the horizon-crossing moment 
but not at the horizon-crossing moment, 
and the coupling strength changes smoothly in time, 
the effect is weaker: for a
detector whose coupling vanishes in any open interval containing the
horizon-crossing moment, the difference from the Minkowski vacuum
response dies off at large $|\omega|$ faster than any inverse power of~$\omega$. 

Our construction of the Rindler firewall state $\hat\rho_{\text{FW}}$
relied on the fact that the right-moving and left-moving components of
a massless field are decoupled in $1+1$ dimensions.  (Related
consequences of this decoupling for past-future correlations have been
investigated
in~\cite{olson-ralph:entanglement,olson-ralph:extraction}.)
$\hat\rho_{\text{FW}}$~is not Hadamard at the Rindler horizon, and we
found that the Wightman function of $\hat\rho_{\text{FW}}$ contains a
heightened version of the $(1+1)$-dimensional infrared ambiguity. 
In particular we found that the response of the derivative-coupling detector is 
ambiguous by an additive Lorentz-noninvariant constant, 
even though this detector is free from
infrared ambiguities in Hadamard states~\cite{Juarez-Aubry:2014jba}. 
It could be interesting to
investigate whether such ambiguities are present for the
derivative-coupling detector in firewall-type states in which a
severing of correlations evolves from an initially regular state by
some dynamical mechanism.

We emphasise that $\hat\rho_{\text{FW}}$ is undoubtedly singular at
the Rindler horizon, as seen from the non-Hadamard form of the Wightman function, 
and from the way in which the detector's
response hinges on the coupling strength at the 
horizon-crossing moment. 
$\hat\rho_{\text{FW}}$~is hence
qualitatively different from an evaporating $(1+1)$-dimensional black
hole in the CGHS model, where the outcome is a long-lived
remnant~\cite{Almheiri:2013wka}, and from a $(1+1)$-dimensional
moving-mirror system that models a remnant~\cite{Hotta:2013clt}.  We
have not attempted to characterise the singularity in
$\hat\rho_{\text{FW}}$ in terms of a stress-energy tensor, or by other
means that might indicate how the spacetime responds to the
singularity when allowed to become dynamical.  However, our main
observation is that when the spacetime is assumed to be unaffected by
the singularity in~$\hat\rho_{\text{FW}}$, the response of the
detector that falls across the horizon is, while sudden, nevertheless
finite.

Our UDW detector had two internal states.  If the detector's internal
Hilbert space is generalised to that of a harmonic oscillator, it
would be usual to take $\mu$ in \eqref{eq:Hint-derivative} to be the
oscillator's position operator, $\mu(\tau) = e^{i\Omega \tau}
d^\dagger + e^{-i\Omega \tau} d$, where $\Omega>0$ is the oscillator's
angular frequency and $(d,d^\dagger)$ are the annihilation and
creation operator pair \cite{Raine:1991kc,Raval:1995mb,Wang:2013lex}.
For the non-derivative detector in $3+1$ dimensions, 
this choice for $\mu$ models the $\bm
p \, \cdot \bm A$ term by which an atomic electron couples to the
quantised electromagnetic field when there is no angular momentum
exchange \cite{MartinMartinez:2012th,alhambra}.  With this choice,
$\mu$ has nonvanishing matrix elements only between neighbouring
energy eigenstates, and the only nonvanishing first-order transition
probabilities from detector state $|n\rangle_D$ are to detector states
$|n+1\rangle_D$ and $|n-1\rangle_D$, given by our formulas with
$\omega = \pm\Omega$.  The conclusion about a finite detector response
on crossing the firewall hence still holds. 
If however $\mu$  
were chosen to have matrix elements of equal magnitude between each
pair of the harmonic oscillator eigenstates, the sum of the first-order
transition probabilities from state $|n\rangle_D$ to all other states
would diverge for the derivative-coupling detector, 
because of the leading term proportional to $\ln(|\Delta n|)$ 
at large~$\Delta n$, 
but be still finite for the non-derivative detector, 
because the leading term is only proportional to 
${(\Delta n)}^{-2}\ln(|\Delta n|)$. 

We considered also a generalisation of $\hat\rho_{\text{FW}}$ 
in which excitations are added behind the Rindler horizon 
in a way that has been 
suggested \cite{braunstein-private} to model the 
energetic curtain of~\cite{braunstein-et-al}. 
We found that the response is qualitatively 
similar to that in $\hat\rho_{\text{FW}}$ 
but can be made arbitrarily large by 
increasing the temperature-like parameter that 
characterises the added excitations. 

Finally, recall that the short-distance behaviour of the
Wightman function becomes more singular as the spacetime dimension
increases. One may hence expect an UDW detector in dimensions higher than
$1+1$ to react to a firewall more violently~\cite{hodgkinson-louko}.
However, the short-distance behaviour of the derivative-coupling detector in 
$1+1$ dimensions is similar
to that of the non-derivative detector in $3+1$ dimensions 
\cite{Juarez-Aubry:2014jba,satz,louko-satz-curved}. This suggests
that our results for the $1+1$ derivative-coupling UDW detector 
may faithfully reflect the response of a non-derivative UDW 
detector that crosses a $(3+1)$-dimensional firewall.

\section*{Acknowledgments}

I thank Don Marolf for asking how a detector responds 
in the state $\hat\rho_{\text{FW}}$ and for helpful correspondence, 
and Sam Braunstein for asking how a detector responds in the state
$\hat\rho_{\text{EC},T}$ considered in Section~\ref{sec:curtain}. 
I~thank Doyeol Ahn and Paul Nation for the invitation to present 
an early version of this work 
at the meeting RQI North 2014, Seoul, Korea, 30 June -- 3 July 2014, and 
several participants, including Eric Brown, Nick Menicucci, 
Don Page and Bill Unruh, for useful comments. 
I~thank an anonymous referee for helpful suggestions. 
This work was supported in part by STFC 
(Theory Consolidated Grant ST/J000388/1).


\appendix

\section{Asymptotics at large $|\omega|$\label{app:asymptotics}}

In this appendix we verify the asymptotic large $|\omega|$ 
expressions 
\eqref{eq:Fboth-0M-inert-large}, 
\eqref{eq:deltaFboth-largeomega}
and~\eqref{eq:deltaFboth-deg-largeomega}. We assume $\omega\ne0$, and 
we denote by $O^{\infty}(\omega^{-1})$ 
a quantity that vanishes faster than any inverse power of 
$\omega$ as $|\omega| \to \infty$.

\subsection{Minkowski vacuum response}

Consider ${}^{\text{in}}\mathcal{F}^{(1)}_{|0_M\rangle}$
\eqref{eq:F1-0M-inert}. Repeated integration by parts, 
integrating the trigonometric factor~\cite{wong}, 
shows that the second term in \eqref{eq:F1-0M-inert} 
is~$O^{\infty}(\omega^{-1})$. 
This gives \eqref{eq:F1-0M-inert-large} in the main text. 

Consider then ${}^{\text{in}}\mathcal{F}^{(0)}_{|0_M\rangle}$
\eqref{eq:F0-0M-inert}. We write 
\begin{subequations}
\begin{align}
{}^{\text{in}}\mathcal{F}^{(0)}_{|0_M\rangle}(\omega) 
&= {}^{\text{in}}\mathcal{F}^{(0)}_1(\omega)
+ 
{}^{\text{in}}\mathcal{F}^{(0)}_2(\omega)
\ , 
\label{eq:inF0-split-def}
\\[1ex]
{}^{\text{in}}\mathcal{F}^{(0)}_1(\omega)
&= - \frac12 
\int^{\infty}_{0} 
ds \sin(\omega s) \, H(s) 
\ , 
\label{eq:inF0split1}
\\[1ex]
{}^{\text{in}}\mathcal{F}^{(0)}_2(\omega)
&= 
- \frac{1}{\pi} 
\int^{\infty}_{0} 
ds \cos(\omega s) \ln(m_0 s) \, H(s) 
\ , 
\label{eq:inF0split2}
\end{align}
\end{subequations}
where $H(s) := \int_{-\infty}^{\infty} d u \, \chi(u)\chi(u-s)$. 
$H$~is a smooth function of compact support, 
it is even, 
and integration by parts shows that 
$H^{(2k)}(0) = {(-1)}^k \int_{-\infty}^{\infty} d u 
\left[\chi^{(k)}(u)\right]^2$ for $k = 0, 1, 2,\ldots$. 

For ${}^{\text{in}}\mathcal{F}^{(0)}_1$, 
repeated integration by parts in \eqref{eq:inF0split1} gives 
\begin{align}
{}^{\text{in}}\mathcal{F}^{(0)}_1(\omega) = 
- \frac12 
\sum_{r=0}^k {(-1)}^{r}\frac{H^{(2r)}(0)}{\omega^{2r+1}}
+ O \! \left(\frac{1}{\omega^{2k+3}}\right)
\ , 
\hspace{2ex}
k = 0, 1, 2, \ldots
\ . 
\label{eq:inF0-split1-f1}
\end{align}

For ${}^{\text{in}}\mathcal{F}^{(0)}_2$, integrating 
\eqref{eq:inF0split2} by parts twice gives 
\begin{align}
{}^{\text{in}}\mathcal{F}^{(0)}_2(\omega) 
& = 
\frac{1}{\pi \omega}
\int_0^{\infty} ds \, \frac{\sin(\omega s)}{s} \, H(s) 
+ 
\frac{1}{\pi \omega^2}
\int_0^{\infty} ds \cos(\omega s) \, \frac{H'(s)}{s}  
\notag
\\[1ex]
& \hspace{3ex}
+ 
\frac{1}{\pi \omega^2}
\int_0^{\infty} ds \cos(\omega s) \ln(m_0 s) \, H''(s) 
\ . 
\label{eq:inF0-split2-t1}
\end{align}
In the first term in \eqref{eq:inF0-split2-t1} we write $H(s) = H(0) +
\bigl[H(s) - H(0)\bigr]$, we use in the part proportional to $H(0)$
the identity $\int_0^\infty dx \, x^{-1} \sin x = \pi/2$, and we
estimate the remainder by repeated integration by parts, finding that
this term equals $\tfrac12 H(0)/|\omega| + O^{\infty}(\omega^{-1})$.
The second term in \eqref{eq:inF0-split2-t1} is
$O^{\infty}(\omega^{-1})$, again using repeated integration by parts.
The last term in \eqref{eq:inF0-split2-t1} has the same form as
\eqref{eq:inF0split2} but with $H \to H''$ and an overall
factor~$-1/\omega^2$. Proceeding recursively, we hence obtain
\begin{align}
{}^{\text{in}}\mathcal{F}^{(0)}_2(\omega) = 
\frac12 
\sum_{r=0}^k {(-1)}^{r}\frac{H^{(2r)}(0)}{{|\omega|}^{2r+1}}
+ O \! \left(\frac{1}{{|\omega|}^{2k+3}}\right)
\ , 
\hspace{2ex}
k = 0, 1, 2, \ldots
\ . 
\label{eq:inF0-split2-f1}
\end{align}

Substituting \eqref{eq:inF0-split1-f1}
and \eqref{eq:inF0-split2-f1} into~\eqref{eq:inF0-split-def}, 
and using the values of $H^{(2k)}(0)$
found above, gives \eqref{eq:F0-0M-inert-large} in the main text.

\subsection{Firewall response}

Consider $\Delta \mathcal{F}^{(1)}$~\eqref{eq:DeltaF1-nonzero}.  The
large $|\omega|$ expansion of the second term can be obtained by
repeated integration by parts, integrating the trigonometric
term~\cite{wong}.  When $\chi(0)\ne0$, the leading terms are shown
in~\eqref{eq:deltaF1-largeomega}. When $\chi(0)=0$, it follows from
the non-negativity of $\chi$ that $\chi'(0)=0$, and the expansion
starts as shown in~\eqref{eq:deltaF1-deg-largeomega}.

Consider then $\Delta \mathcal{F}^{(0)}$~\eqref{eq:DeltaF0}. We write
\begin{subequations}
\begin{align}
\Delta \mathcal{F}^{(0)}(\omega) 
&= \Delta \mathcal{F}^{(0)}_1(\omega) 
+ 
\Delta \mathcal{F}^{(0)}_2(\omega) 
\ , 
\label{eq:DeltaF0-split-def}
\\[1ex]
\Delta \mathcal{F}^{(0)}_1(\omega) 
&= \frac14
\int^{\infty}_{0} 
ds \sin(\omega s) \, G(s) 
\ , 
\label{eq:DeltaF0split1}
\\[1ex]
\Delta \mathcal{F}^{(0)}_2(\omega) 
&= 
\frac{1}{2\pi} 
\int^{\infty}_{0} 
ds \cos(\omega s) \ln(\mtilde s) \, G(s) 
\ , 
\label{eq:DeltaF0split2}
\end{align}
\end{subequations}
where $\mtilde:= m_0 e^{\eta\beta}$ and $G(s) := \int_0^s d u \,
\chi(u)\chi(u-s)$. 
$G$~is a smooth function of compact support, it is
odd, and we have $G'(0) = {[\chi(0)]}^2$, $G^{(3)}(0) = 2 \chi(0)
\chi''(0) - {[\chi'(0)]}^2$ and $G^{(5)}(0) = 2 \chi(0) \chi^{(4)}(0)
- 2 \chi'(0) \chi^{(3)}(0) + {[\chi''(0)]}^2$.

For $\Delta \mathcal{F}^{(0)}_1$,  
repeated integration by parts in \eqref{eq:DeltaF0split1}
gives $\Delta \mathcal{F}^{(0)}_1(\omega) = O^{\infty}(\omega^{-1})$. 

For $\Delta \mathcal{F}^{(0)}_2$, integration by parts in
\eqref{eq:DeltaF0split2} gives
\begin{align}
\Delta \mathcal{F}^{(0)}_2(\omega) 
&= 
- \frac{1}{2\pi \omega} 
\int^{\infty}_{0} 
ds \sin(\omega s)\, \frac{G(s)}{s}
\; - \; \frac{1}{2\pi \omega} 
\int^{\infty}_{0} 
ds \sin(\omega s) \ln(\mtilde s) \, G'(s) 
\ .  
\label{eq:app:s21}
\end{align}
To handle the second term in~\eqref{eq:app:s21}, 
we introduce a cutoff $\epsilon>0$ and observe that 
\begin{align}
\omega \int^{\infty}_\epsilon 
ds \sin(\omega s) \ln(\mtilde s) \, G'(s) 
&= \cos(\omega \epsilon) \ln(\mtilde \epsilon) \, G'(\epsilon)
- G'(0) \Ci(\epsilon |\omega|) 
\notag
\\[1ex]
& \hspace{3ex}
+ \int^{\infty}_\epsilon 
ds \cos(\omega s) \, \frac{G'(s) - G'(0)}{s}
\notag
\\[1ex]
& \hspace{3ex}
+ \int^{\infty}_\epsilon 
ds \cos(\omega s) \ln(\mtilde s) \, G''(s) 
\ , 
\label{eq:app:s22}
\end{align}
first integrating by parts and then subtracting and adding $G'(0)
\Ci(|\omega|\epsilon)$, where $\Ci$ is the cosine integral function in
the notation of~\cite{dlmf}. The limit $\epsilon\to0_+$ in
\eqref{eq:app:s22} can be taken using the small argument form of
$\Ci$~\cite{dlmf}, and substituting the result in \eqref{eq:app:s21}
yields
\begin{align}
\Delta \mathcal{F}^{(0)}_2(\omega) 
&= 
- \frac{1}{2\pi \omega} 
\int^{\infty}_{0} 
ds \sin(\omega s)\, \frac{G(s)}{s}
\; + \; \frac{G'(0)}{2\pi \omega^2} 
\ln \bigl( |\omega|e^{\gamma}/\mtilde\bigr)
\notag
\\[1ex]
& \hspace{3ex}
- \frac{1}{2\pi \omega^2} 
\int^{\infty}_0
ds \cos(\omega s) \, \frac{G'(s) - G'(0)}{s}
\notag
\\[1ex]
& \hspace{3ex}
- \frac{1}{2\pi \omega^2} 
\int^{\infty}_{0} 
ds \cos(\omega s) \ln(\mtilde s) \, G''(s) 
\ ,  
\label{eq:app:s23}
\end{align}
where $\gamma$ is Euler's constant. Repeated integration by parts
gives for the first term in \eqref{eq:app:s23} an expansion in inverse
powers of~$\omega^2$, and the same technique shows that the third term
in \eqref{eq:app:s23} is~$O^{\infty}(\omega^{-1})$. We find 
\begin{align}
\Delta \mathcal{F}^{(0)}_2(\omega) 
&= 
- \frac{1}{2\pi \omega^2} 
\int^{\infty}_{0} 
ds \cos(\omega s) \ln(\mtilde s) \, G''(s) 
+ \frac{G'(0)}{2\pi \omega^2} 
\ln \bigl( |\omega|e^{\gamma-1}/\mtilde\bigr)
\notag
\\[1ex]
& \hspace{3ex}
+ \frac{1}{2\pi} 
\sum_{r=2}^k {(-1)}^{r}\frac{G^{(2r-1)}(0)}{{(2r-1)\omega}^{2r}}
+ O \! \left(\frac{1}{{\omega}^{2k+2}}\right)
\ , 
\hspace{2ex}
k = 2, 3, 4, \ldots
\label{eq:app:s24}
\end{align}

Now, the first term in \eqref{eq:app:s24} has the same form as
\eqref{eq:DeltaF0split2} but with $G \to G''$ and an overall
factor~$-1/\omega^2$, and we may proceed with $\Delta
\mathcal{F}^{(0)}_2$ recursively.  Collecting, we find for $\Delta
\mathcal{F}^{(0)}$ the asymptotic large $|\omega|$ expansion
\begin{align}
\Delta \mathcal{F}^{(0)}(\omega) 
&\sim 
\frac{1}{2\pi} 
\ln \bigl( |\omega|e^{\gamma-1}/\mtilde\bigr)
\left(
\frac{G'(0)}{\omega^2} 
- \frac{G^{(3)}(0)}{\omega^4}
+ \frac{G^{(5)}(0)}{\omega^6}
- \frac{G^{(7)}(0)}{\omega^8}
+ \cdots 
\right)
\notag
\\[1ex]
& \hspace{3ex}
+ \frac{1}{2\pi}
\left(\frac{\frac13 G^{(3)}(0)}{\omega^4} 
- \frac{\left(\frac13 + \frac15\right)G^{(5)}(0)}{\omega^6} 
+  \frac{\left(\frac13 + \frac15 + \frac17\right)G^{(7)}(0)}{\omega^8} 
- \cdots 
\right) 
\ . 
\label{eq:app:s25}
\end{align}
Equations \eqref{eq:deltaF0-largeomega} and
\eqref{eq:deltaF0-deg-largeomega} in the main text follow from
\eqref{eq:app:s25} by inserting the values of $G'(0)$, $G^{(3)}(0)$ and
$G^{(5)}(0)$ found above.

\section{Evaluation of 
$\Delta \mathcal{F}^{(0)}$ 
and $\Delta \mathcal{F}^{(1)}$\label{app:evaluation}}

In this appendix we verify formulas 
\eqref{eq:DeltaFboth} 
for 
$\Delta \mathcal{F}^{(0)}$ 
and $\Delta \mathcal{F}^{(1)}$. 
We write $\mtilde:= m_0 e^{\eta\beta}$, 
$Q(\tau) := e^{-i\omega\tau}\chi(\tau)$ and  
$Q'(\tau) := \frac{d}{d\tau}Q(\tau)$.

\subsection{$\Delta \mathcal{F}^{(0)}$}

Starting from \eqref{eq:DeltaF-def} with $p=0$, 
we have 
\begin{align}
\Delta \mathcal{F}^{(0)}(\omega)
&=
\int^{\infty}_{-\infty}\,d\tau'\,\int^{\infty}_{-\infty}\,d\tau''\, 
Q(\tau') \overline{Q(\tau'')} 
\, \Delta \mathcal{W}(\tau',\tau'')
\notag
\\[1ex]
&=
\Realpart
\int^{\infty}_0\,d\tau'\, 
\int^0_{-\infty}\,d\tau''\, 
\left\{ 
{(2\pi)}^{-1}\ln \bigl[\mtilde (\tau'-\tau'') \bigr]
+ \tfrac14 i 
\right\}
Q(\tau') \overline{Q(\tau'')} 
\ , 
\end{align}
using 
\eqref{eq:DeltaW-def} for $\Delta\mathcal{W}$ 
and interchanging the names of $\tau'$ and $\tau''$ 
in the region where originally $\tau'< 0 < \tau''$. 
Writing 
$u := \tau'$ and $\tau'' = u-s$, intechanging the integration order, 
and using $Q(\tau) = e^{-i\omega\tau}\chi(\tau)$, 
we obtain 
\begin{align}
\Delta \mathcal{F}^{(0)}(\omega)
&= 
\int_0^\infty 
\,ds 
\left[ \tfrac14 \sin(\omega s) + {(2\pi)}^{-1} \cos(\omega s) \ln(\mtilde s) \right]
\, \int_0^s du \, 
\chi(u) \chi(u-s) 
\ , 
\end{align}
which is equation \eqref{eq:DeltaF0} in the main text.

\subsection{$\Delta \mathcal{F}^{(1)}$}

Starting from \eqref{eq:DeltaF-def} with $p=1$, 
we have 
\begin{align}
\Delta \mathcal{F}^{(1)}(\omega)
&=
\int^{\infty}_{-\infty}\,d\tau'\,\int^{\infty}_{-\infty}\,d\tau''\, 
Q'(\tau') \overline{Q'(\tau'')} 
\, \Delta \mathcal{W}(\tau',\tau'')
\notag
\\[1ex]
&=
\frac{1}{2\pi}
\Realpart
\int^{\infty}_0\,d\tau'\, 
\int^0_{-\infty}\,d\tau''\, 
\ln \bigl[\mtilde (\tau'-\tau'') \bigr]
\, 
Q'(\tau') \overline{Q'(\tau'')} 
\ , 
\end{align}
first integrating the distributional derivatives by parts, then using
\eqref{eq:DeltaW-def} for $\Delta\mathcal{W}$ and noting that the
contributions from the $\pm \frac18 i$ terms in \eqref{eq:DeltaW-def}
cancel, and finally interchanging the names of $\tau'$ and $\tau''$ 
in the region where originally $\tau'< 0 < \tau''$. 
Writing 
$u := \tau'$ and $\tau'' = u-s$, and intechanging the integration
order, we obtain
\begin{align}
\Delta \mathcal{F}^{(1)}(\omega)
=
\frac{1}{2\pi}
\Realpart
\int^{\infty}_0\,ds \, \ln(\mtilde s) 
\int_0^s\,du \, 
Q'(u) \overline{Q'(u-s)} 
\ . 
\label{eq:app:t00}
\end{align}

Using in \eqref{eq:app:t00} the identity 
\begin{align}
\int_0^s\,du \, 
Q'(u) \overline{Q'(u-s)} 
= \frac{d}{ds}
\left( 
Q(0) \overline{Q(-s)} 
+ \int_0^s\,du \, 
Q(u) \overline{Q'(u-s)} 
\right) 
\ , 
\end{align}
separating the two terms and integrating the second term by parts, we find 
\begin{subequations}
\label{eq:app:t0}
\begin{align} 
\Delta \mathcal{F}^{(1)}(\omega) 
& = 
\Delta \mathcal{F}^{(1)}_1(\omega) + \Delta \mathcal{F}^{(1)}_2(\omega) 
\ , 
\label{eq:app:t0a}
\\[1ex]
\Delta \mathcal{F}^{(1)}_1(\omega) 
&= 
\frac{\chi(0)}{2\pi}
\int^{\infty}_0\,ds \, \ln(\mtilde s) 
\, \frac{d}{ds}
\bigl[ \cos(\omega s) \chi(-s) \bigr] 
\ , 
\label{eq:app:t0b}
\\[1ex]
\Delta \mathcal{F}^{(1)}_2(\omega) 
& = 
- \frac{1}{2\pi}
\Realpart
\int^{\infty}_0 \, 
\frac{ds}{s}
\int_0^s\,du \, 
Q(u) \overline{Q'(u-s)} 
\ . 
\label{eq:app:t0c}
\end{align}
\end{subequations}

Consider first~$\Delta \mathcal{F}^{(1)}_1$~\eqref{eq:app:t0b}. When
$\omega=0$, \eqref{eq:app:t0b} reduces to 
\begin{align}
\Delta \mathcal{F}^{(1)}_1(0)
&=
- \frac{\chi(0)}{2\pi}
\int^{\infty}_0\,ds \, \ln(\mtilde s) 
\, 
\chi'(-s) 
\ . 
\label{eq:app:t11}
\end{align}
When $\omega\ne0$, we introduce a cutoff $\epsilon>0$ and write  
\begin{align}
\int^{\infty}_\epsilon\,ds \, \ln(\mtilde s) \, 
\frac{d}{ds}
\bigl[ \cos(\omega s) \chi(-s) \bigr] 
&= 
- \ln(\mtilde \epsilon) 
\cos(\omega \epsilon) \chi(-\epsilon)
- 
\int^{\infty}_\epsilon
\,
\frac{ds}{s} 
\cos(\omega s) \chi(-s)
\notag
\\[1ex]
&= 
- \ln(\mtilde \epsilon) 
\cos(\omega \epsilon) \chi(-\epsilon)
+ \chi(0) \Ci(|\omega|\epsilon) 
\notag
\\[1ex]
& \hspace{3ex}
+ 
\int^{\infty}_\epsilon
\,
ds 
\cos(\omega s) \, \frac{\bigl[\chi(0) - \chi(-s)\bigr]}{s}  
\ , 
\label{eq:app:t12}
\end{align}
integrating by parts and adding and subtracting $\chi(0)
\Ci(|\omega|\epsilon)$. Using the small argument form of $\Ci$ to take
the limit~\cite{dlmf}, we find 
\begin{align}
\Delta \mathcal{F}^{(1)}_1(\omega)
& =
\frac{{[\chi(0)]}^2}{2\pi}
\ln \bigl( |\omega|e^{\gamma}/\mtilde\bigr)
\ + \  
\frac{\chi(0)}{2\pi}
\int_0^\infty 
ds
\, \cos(\omega s) 
\, 
\frac{\bigl[\chi(0)
- \chi(-s) \bigr]}{s} 
\ , 
\label{eq:app:t13}
\end{align} 
where $\gamma$ is Euler's constant. 

Consider then~$\Delta \mathcal{F}^{(1)}_2$~\eqref{eq:app:t0c}. 
Using in \eqref{eq:app:t0c} the identity 
\begin{align}
- \Realpart 
\int_0^s\,du \, 
Q(u) \overline{Q'(u-s)} 
= - \cos(\omega s) \chi(0) \chi(s)
+\frac{d}{ds}
\int_0^s du \, 
\cos(\omega s) \chi(u) \chi(u-s) 
\label{eq:app:ident55}
\end{align}
and integrating the second term in 
\eqref{eq:app:ident55} by parts, we find 
\begin{align}
\Delta \mathcal{F}^{(1)}_2(\omega) 
&= 
\frac{1}{2\pi}\lim_{\epsilon\to0_+}
\left\{- \frac{1}{\epsilon} 
\int_0^\epsilon du \, 
\chi(u) \chi(u-\epsilon) 
\right. 
\notag
\\[1ex]
& \hspace*{3ex}
\left. 
+ \int^{\infty}_\epsilon 
ds 
\cos(\omega s) 
\left[ 
- \frac{\chi(0) \chi(s)}{s}
+ \frac{1}{s^2}
\int_0^s du \, 
\chi(u) \chi(u-s) 
\right]
\right\} 
\notag
\\[1ex]
& = 
-\frac{{[\chi(0)]}^2}{2\pi} 
+ 
\frac{1}{2\pi}
\int^{\infty}_0
ds 
\cos(\omega s) 
\left[ 
- \frac{\chi(0) \chi(s)}{s}
+ \frac{1}{s^2}
\int_0^s du \, 
\chi(u) \chi(u-s) 
\right] 
\ . 
\label{eq:app:t21}
\end{align}

For $\omega\ne0$, combining 
\eqref{eq:app:t13} and \eqref{eq:app:t21}
gives \eqref{eq:DeltaF1-nonzero} in the main text. 

For $\omega=0$, 
we set $\omega=0$ in \eqref{eq:app:t21}, 
we add and subtract under the $s$-integral the term 
$\chi(0) \bigl[ \chi(-s) - \chi(s)\bigr] {(2s)}^{-1}$, 
and we integrate the added term by parts. 
Combining with \eqref{eq:app:t11} gives \eqref{eq:DeltaF1-zero} 
in the main text.

\end{document}